\documentstyle[12pt]{article}
\begin{document}
\begin{titlepage}
\nopagebreak
\begin{flushright}

LPTENS--91/19\\
                July,~1991

\end{flushright}

\vfill
\begin{center}
{\large\bf
Classical $W_n$-Symmetry\\
and Grassmannian Manifold}

\vfill
{\bf Y.~Matsuo}\\
Laboratoire de Physique Th\'eorique\\
Ecole Normale Sup\'erieure\footnote{Unit\'e propre du
Centre National de la Recherche Scientifique,
associ\'ee \`a l'Ecole Normale Sup\'erieure et \`a l'Universit\'e
de Paris-Sud},\\
24 rue Lhomond, 75231 Paris CEDEX 05, ~France\\

\end{center}
\vfill

\begin{abstract}

The classical $W_n$-symmetry is globally parametrized by
the Grassmannian manifold which is described by the
nonrelativistic fermions.  We give the bosonization rule which defines
the natural higher coordinates system to describe the $W_n$-geometry.
The generators of the $W_n$-algebra can be obtained from
the $\tau$-function by using the vertex operator.

\end{abstract}
\vfill

\end{titlepage}

\section{Introduction}
Recently, a lot of attentions have been attracted to
the investigation of the geometrical structure of
the so-called $W_n$-gravity or the
$W_n$-geometry.\cite{Landau}-\cite{Saclay}
This problem is challenging because it provides the
natural extension of the two dimensional gravity
and it will possibly be related to the higher dimensional
field theories.\cite{SDYM}
Although there are many works which discussed on
such geometry, it is still far from  the
complete understanding
because
of its nonlinearity.

In this paper, we remark the relation between the
global structure of the classical $W_n$-symmetry
and the  Grassmannian manifold.
It is known\cite{GY} that the $W_n$-symmetry is closely related to
the KdV equations.\cite{Sato}\cite{DS}
Both of them are described as the deformation of rank $n$ differential
equations and the space of differential operators is
parametrized by the Grassmannian manifold.
In  the KdV equations
(or KP hierarchy ), the nonlinear equations
are mapped to linear motions on the Universal Grassmannian
Manifold (UGM).\cite{Sato}
Our idea is that non-linear structure of $W_n$ symmetry
will be simplified also in the language of the Grassmannian.
Although we have not fully completed this program, the
mere introduction of the Grassmannian seems to provide many
clues to understand the geometry.
It is known that the natural tool to describe
the Grassmannian manifold is the free fermion theory.
Unlike the KP hierarchy theory,
our fermions turns out to be  non-relativistic. It reflects the fact that
the particle number is always finite.
We show that the deformation of $n$-th order differential operators
 can be linearly described
by the bosonization.  We prove that the bosonic degree of
freedom is finite dimensional ( say $t_1,\cdots, t_n$ ) and the first
one ( $t_1$ ) corresponds to the holomorphic coordinate of the
Riemann surface.

The most important merit to introduce such higher coordinates is that
$W_n$-algebra generators can then be obtained from a single generator,
i.e. the $\tau$-function.  In this sense, the situation becomes similar to
that of the super Virasoro symmetry with the supercoordinate.

\section{Classical $W_n$-symmetry}
In various contexts,  the classical $W_n$-symmetry is realized by
the deformation of $n$-th order differential
equation,\cite{Landau}\cite{BG}\cite{Sotokov}\cite{Saclay}

\begin{eqnarray}
{\cal L}_n f(t) & = & 0,\label{Diff:eq}\\
{\cal L}_n & = & \partial^n + u_1(t)\partial^{n-1} +
u_2(t)\partial^{n-2}+ \cdots
u_n(t).\label{Ln}
\end{eqnarray}

We call the coefficient functions $u_s(t)$  gauge potentials
in the following.
In order to get the $W_n$ symmetry, we need to require the constraint
for the first gauge potential,

\begin{equation}
\label{u1:const}
u_1(t) \equiv 0.
\end{equation}

The  deformation
is generated by a pair of
differential operators $X$ and $Y$\footnote{
We use the notation of the ref. \cite{Saclay}}.
We need to choose them such that ${\cal L}_n$ keeps its
form (\ref{Ln},\ref{u1:const})
under the variation of the following type,

\begin{eqnarray}
\delta_X {\cal L}_n & \equiv & Y {\cal L}_n - {\cal L}_n X
\label{var:XY}\\
\label{var:Ln}
   &  = & \delta_X u_2\partial^{n-2} + \cdots + \delta_X u_n,\\
\delta_X f & = & X f(t) \label{var:psi}
\end{eqnarray}

The relation between the operators $X$, $Y$ is important.
{}From eq.(\ref{var:XY}), $Y$ is formally given by
\begin{eqnarray}
Y & = & {\cal L}_nX{\cal L}_n^{-1} + \delta{\cal L}_n{\cal L}_n^{-1}
\nonumber\\
  &  =  &  ({\cal L}_n X {\cal L}_n^{-1})^+.  \label{Y=F(X)}
\end{eqnarray}
Here we used the projection of the pseudo differential operator
into positive parts $(~)^+$.  Explicitly, if $O = \sum_{\ell=0}^\infty
o_\ell \partial^{n-\ell}$, we get $(O)^+ = \sum_{\ell = 0}^n o_\ell
\partial^{n-\ell}$.  We define also $(O)^- = O-(O)^+$.
If we ignore the constraint (\ref{u1:const}), we can start from any
differential operator $X$ to get the deformation of ${\cal L}_n$,
\begin{equation}
\label{var1:Ln}
\delta_X{\cal L}_n = {\cal B}_X{\cal L}_n, \qquad
{\cal B}_X = - ({\cal L}_n X {\cal L}_n^{-1})^-.
\end{equation}
We remark that although the deformation equation
(\ref{var1:Ln}) reminds us of that of the KP hierarchy,\cite{Sato}
there are some important differences.
In the KP hierarchy, we define the $0$-th order pseudo-differential
operator (PSD) $S= 1 + s_1(t)\partial^{-1} +s_2(t)\partial^{-2}+\cdots$
through ${\cal L}_n = S\partial^nS^{-1}$.  Such operator  can be
determined up to right multiplication of constant $0$-th order
PSD $C$, $S\rightarrow SC$.  Then ``time evolution" of the system
is governed by $\frac{\partial}{\partial t_\ell} S = A_\ell S$
with $A_\ell = - (S\partial^\ell S^{-1})^-.$
These time evolutions are then interpreted as  linear motions
on the Universal Grassmannian Manifold which is directly related
to $S$.  This formulation has a merit that we can discuss
the deformation of the differential operators of any order in the same
framework.
However,  the relation between $s_\ell$ and gauge potentials
is nonlinear.  Moreover, the Poisson bracket structure of
${\cal L}_n$ is written in terms of the variation (\ref{var1:Ln}).
If we the inner product  $(A,{\it L}_n) = \int dt \sum_{\ell =1}^n a_\ell(t)
u_\ell(t)$ for $A = \sum_{\ell =1}^n \partial^{\ell -1-n}a_\ell(t)$,
the second Poisson bracket\cite{DS} is written by,
\begin{equation}
\{ (A,{\cal L}_n), (B,{\cal L}_n)\} = -(B, ({\cal L}_n A)^-{\cal L}_n).
\end{equation}
The RHS of this equation is the variation of ${\cal L}_n$ with
$X = (A{\cal L}_n)^+$.  This gives the classical definition of
the $W_n$ algebra.
Hence in order to understand its geometrical structure,
we had better introduce the Grassmanian which
is directly related to ${\cal L}_n$ itself.

\section{Grassmannian manifold}
The correspondence between $n$-th order
differential equation and the Grassmannian is more or less
known through the essential role of the Wronskian.
In our paper, we make it in more explicit form
and develop the suitable machinery to describe it.

Define a Hilbert space ${\cal H}$  as the space of functions
which is regular at $t=0$, i.e. $f\in {\cal H}$ if
it has the Taylor expansion,
\begin{equation}
\label{f}
f(t) =
\sum_{n=0}^\infty \frac{f_n}{n!}t^n.
\end{equation}
Suppose all the coefficients in ${\cal L}_n$ are regular at
$t=0$\footnote{ If we allow that the gauge potentials might have
singularities at the origin $t=0$, we need to extend the
Hilbert space to {\bf C}[[$\log t$]] instead of {\bf C}[[$t$]].
We must modify our discussion in the following if we need to
discuss this generalized situation.}.
  Then  the kernel the differential operator
(\ref{Diff:eq}) determines an $n$-dimensional subspace of ${\cal H}$.
Denote
the Grassmannian manifold which consists of $n$-dimensional subspaces
of ${\cal H}$ by $Gr^{(n)}$.
The correspondence discussed above shows that
there is a mapping from the space of $n$-th order differential
operators into the Grassmanian $Gr^{(n)}$.
It is important to notice that there is the inverse mapping.
Let $\Omega$ be a point of $Gr^{(n)}$.  Pick up a basis
of $\Omega$, $\{ f_1, \cdots, f_n\}$.
We can get $n$-th order differential equation,
\begin{equation}
{\cal L}_n(\Omega)f
= \tau_n(\Omega)^{-1}\left| \begin{array}{cccc}
f & f_1 & \cdots & f_n\\
\partial f & \partial f_1& \cdots & \partial f_n\\
\vdots & \vdots & ~      & \vdots\\
\partial^n f & \partial^n f_1 & \cdots & \partial^n f_n
\end{array} \right|
\label{Ln:omega}
\end{equation}
where,
${\cal L}_n(\Omega) \equiv \partial^n + u_1(\Omega)\partial^{n-1}
+ \cdots + u_n(\Omega)$ and
\begin{equation}
\label{Dn}
\tau_n(\Omega) \equiv \left| \begin{array}{ccc}
f_1 & \cdots & f_n \\
\vdots & ~ & \vdots \\
\partial^{n-1} f_1 & \cdots & \partial^{n-1} f_n \end{array} \right|.
\end{equation}
In this way, one observe the bijection between
the Grassmanian $Gr^{(n)}$ and the space of the $n$-th order differential
operators ( without $u_1$ constraint (\ref{u1:const})).
In a sense, the functions $\{u_1(\Omega), \cdots , u_n(\Omega)\},$
can be regarded as inhomogeneous coordinates of $Gr^{(n)}$.

Now let us come back to the classical $W_n$-symmetry.
It is defined as the deformations of ${\cal L}_n$.
Since the Grassmannian $Gr^{(n)}$ parametrizes the
family of differential
operators,
$W_n$-algebra generators can be regarded
as the elements of the tangent space
of the Grassmannian, $T_*Gr^{(n)}$\footnote{
Precisely, we need to impose the constraint
to get the $W_n$ symmetry.  It will be discussed in the last section.}.
In this sense, the
Grassmannian gives the global structure of the $W_n$-algebra.
The merit to lay emphasis on  the Grassmannian is that it is
a homogenious space and basically a linear object.
As we are going to discuss, we can treat it neatly  through
the free fermion operator formalism and its bosonization.

\section{Free fermion formalism}
As in the KP hierarchy, the free fermion operator
formalism provides a systematic and powerful tool to analyze the
structure of the Grassmannian.
In order to apply to our case, however, we need to make a slight
modification.  It is due to the fact that
the Wronskians which appeared in the previous section had
finite ranks.  Correspondingly, we need to deal with the
states with finite particle number.
This situation is similar to the free fermion theory
of the finite $N$ matrix models.\cite{RU}
In fact, the nonrelativistic fermion theory used there can
be also applied to our case.

Let us define the free fermion operators and the Fock vacuum
as in the matrix model.
\begin{eqnarray}
\{\psi_n, \psi_m\} & = & \{\psi^*_n, \psi^*_m\} = 0,\\
\{\psi_n, \psi^*_m\} & = & \delta_{n,m},\qquad (~n,m~=~0,~1,~\cdots)\\
\psi^*_n|\emptyset> & = & 0 \qquad  \forall n,\\
<\emptyset|\psi_n & = & 0  \qquad \forall n.
\end{eqnarray}
Note that we use the semi-infinite indices $n=0,1,2,\cdots$ for
the fermion operators.
The vacuum states $|\emptyset>$ and $<\emptyset|$ correspond to the no
particle states.  The $n$--particle ground state can be created from
them in the standard way,
\begin{eqnarray}
|n> & = & \psi_{n-1}\psi_{n-2}\cdots\psi_0|\emptyset>,\\
<n| & = & <\emptyset|\psi^*_0\psi^*_1\cdots\psi^*_{n-1}.
\end{eqnarray}
The $U(1)$ current operators,
\begin{equation}
J_n = \sum_{s=0}^\infty \psi_{n+s}\psi^*_s,
\end{equation}
shall play the role of the ``Hamiltonian" as it does in the
KP hierarchy.

We represent the Hilbert space ${\cal H}$ by the ``one-particle
state" in the fermion Fock space.  For each $f(t)=\sum_n \frac{f_n}{n!}
t^n \in {\cal H}$, define $\psi^*_f \equiv \sum_nf_n\psi^*_n$.
We can show,
\begin{equation}
f(t) = <\emptyset|\psi^*_f \exp(J_1t)|1>.
\end{equation}
More generally, we can represent $Gr^{(n)}$ in our Fock space.
For each $\Omega\in Gr^{(n)}$, choose a frame $\{f_1,\cdots, f_n\}$.
It is mapped into the $n$-particle states $<\Omega|\equiv
<\emptyset|\psi^*_{f_1}\cdots\psi^*_{f_n}$.
This procedure has an ambiguity up to the multiplication of the
constant number which depends on the choice of the frame.
This is the standard Pl\"ucker embedding\footnote{
Conversely, generic points of the $n$-particle states
do not correspond to the Grassmannian.
The  Grassmannian points
are characterized by the Hirota's bilinear identity,
$\sum_{s=0}^\infty <\Omega|\psi_s\bigotimes<\Omega|\psi^*_s = 0$.}.
It is easily proved that the determinant $\tau_n(\Omega)$
(12) can be represented as the vacuum expectation value
(VEV) of $n$-particle states,
\begin{equation}
\tau_n(\Omega) = <\Omega|\exp(J_1t)|n>.
\end{equation}
The differential equation is now written simply as
$<\Omega|\psi^*_f = 0$.
Instead of using this form, it is more convenient to
introduce  following
``generating functional" for the gauge potentials,
\begin{eqnarray}
W_\Omega(\zeta) & = & \sum_{s=0}^n\zeta^su_s(\Omega)\tau_n\nonumber\\
    & = & <\Omega|\exp(J_1t)\psi^*(\zeta)|n+1>.
\end{eqnarray}
with $\psi^*(\zeta) = \sum_{n=0}^\infty \psi^*_n\zeta^n.$
In order to prove these identities, we used a lemma,
$\partial^nf(t) = <\emptyset|\psi_f \exp(tJ_1)\psi_n|\emptyset>$.
In the terminology of the KP hierarchy, $\tau_n$ and
$W_\Omega(\zeta)$ correspond to the $\tau$-function and the Baker-Akhiezer
function respectively.
More generally,  we define the generating functional of the
character functions,
\begin{equation}
W_\Omega(\zeta_1,\cdots,\zeta_m)  =
<\Omega|\exp(J_1t)\psi^*(\zeta_1)\cdots\psi^*(\zeta_m)|n+m>.
\end{equation}

\section{Bosonization}
We would like to discuss how the deformation of the
differential equation looks like in our fermion formalism.
The natural transformation which acts on the Grassmannian is
the general linear transformation of ${\cal H}$, i.e.
so called $w_{1+\infty}$ algebra.\cite{Winf}
In the free fermion language, the transformation is generated by
the fermion bilinear,
\begin{equation}
w_{n,m} = \sum_{s=m}^\infty \frac{s!}{(s-m)!}\psi_{s+n-m}\psi^*_s.
\end{equation}
The commutation relation with $\psi^*(\zeta)$ illuminates the
origin of this operator,
\begin{equation}
[w_{n,m}, \psi^*(\zeta)] = -\zeta^n\frac{\partial^m}{\partial \zeta^m}
\psi^*(\zeta).
\end{equation}
When it applied to the one particle state
it induces the $w_{1+\infty}$ transformation,
\begin{equation}
\label{deltaf}
<\emptyset|\psi^*_fw_{n,m}\exp(tJ_1)|1> = t^m\frac{\partial^n}{
\partial t^n}f(t).
\end{equation}
We should remark that the indices $n$ and $m$ are interchanged.
It is because  the Hamilonian $\exp(tJ_1)$ actually plays the
role of Laplace transformation here.  Similar phenomena also
happened in the finite N matrix models.

Now the variation of the $n$-particle bra state $<\Omega|$ is simply given
by,
$\delta_{n,m}<\Omega|= <\Omega|w_{n,m}$.
We need to rewrite its action in terms of the bosonic variable.
However, in order to describe the multi-particle states,
the single holomorphic coordinate $t$ is not enough.
As in the KP hierarchy, we need to introduce the infinite number of
time coordinates $t_n$ ($n=0,1,2,\cdots$) through the modification
of the Hamiltonian,
\begin{equation}
\exp(tJ_1) \rightarrow \exp(\sum_{\ell=0}^\infty t_\ell J_\ell)
\equiv \exp ( H([t]) ).
\end{equation}
Although we shall show later that
only $n$ coordinates are enough
to describe $n$-particle states,
it is convenient to keep all of them for a while.

The bosonization of the non-relativistic fermion
in the context of the matrix model was studied previously by
Itoyama and current author.\cite{IM}  We can copy it to apply
to this case.
Define
$w_n(\zeta) \equiv \sum_{m=0}^\infty w_{m,n}\zeta^{-m-1}.$
The key identity\footnote{
We need a ``translation" from identities obtained in \cite{IM}.
It is important to notice that, $$det(\lambda_i^j)_{i,j = 1,\cdots,n} =
<\emptyset|\psi^*(\lambda_1)\cdots\psi^*(\lambda_n)|n>,$$
$$<\emptyset|\psi^*(\lambda_1)\cdots\psi^*(\lambda_n)w_{p,q}
= \sum_{i=1}^n\lambda_i^p\frac{\partial^q}{\partial \lambda_i^q}
<\emptyset|\psi^*(\lambda_1)\cdots \psi^*(\lambda_n).$$} is,
\begin{equation}
w_n(\zeta)|k> =
        \frac{1}{n+1}Q_{n+1}(w_0(\zeta))|k>.
\end{equation}
with
\begin{equation}
Q_n(f) = (\partial_\zeta + f(\zeta))^n\cdot 1
= e^{-F(\zeta)}\partial^n e^{F(\zeta)}, \quad (\partial F \equiv f).
\end{equation}
We shall prove later that
$Q_\ell(w_0(\zeta))|k> =0$ identically if $\ell>k$.
We also use the following lemma,
\begin{equation}
w_n(\zeta)\exp(H[t]) = \exp(H([t]))\left(\sum_{\ell = 0}^n
{}~_nC_\ell Q_{n-\ell}(j(\zeta))w_\ell(\zeta)\right)_-,
\end{equation}
where $(~)_-$ is the projection to the negative power of $\zeta$.
In the following, we use the notaions,
 $j(\zeta) = \sum_{\ell=1}^\infty \ell t_\ell \zeta^{\ell-1}$.
and   $\frac{d}{dj} =
\sum_{\ell = 0}^\infty \zeta^{-\ell-1}\frac{\partial}{\partial t_\ell}$.

Combining these identities, we get the transformations of the
$\tau$-function in the bosonized form,
\begin{equation}
\label{var:tau}
\delta_s(\zeta)\tau_n = \tilde{\delta}_s(\zeta) \tau_n,
\end{equation}
with
\begin{equation}
\tilde{\delta}_s(\zeta) = \sum_{\ell =1}^{s+1}~_{s+1}C_\ell
\left(Q_{s+1-\ell}(j(\zeta))
Q_{\ell}(\frac{d}{dj}(\zeta))\right)_-.
\end{equation}
Here, we use the notation $(~)_-$ which means the projection to
the negative powers of $\zeta$.

Eq.(\ref{var:tau}) gives us very important corollary.
Since every $w_{1+\infty}$ generators are bosonized,
we can get more direct ``bosonization" rules.
The most useful one is that
\begin{equation}
\exp(H([t]))\psi^*(\zeta)|n+1>  =  V^*_n(\zeta)\exp(H([t]))|n>,
\end{equation}
with,
\begin{equation}
V^*_n(\zeta)  =
\zeta^n \exp\left(-\sum_{\ell=1}^\infty \frac{\zeta^{-\ell  }}{\ell}
\frac{\partial}{\partial t_\ell}\right).
\end{equation}
This relation explicitly shows that we can get $W_\Omega(\zeta)$
out of $\tau$-function,
$W_\Omega(\zeta) =
                V^*_n(\zeta)\tau_n([t])$.
It seems rather nontrivial that the RHS is a $n$-th polynomial in terms
of $\zeta$.  We shall explicitly prove it later.
Anyway, the importance of this formula is that it implies that
the $W_n$ charges can be obtained from $\tau$-function in a simple way.
In this sense, we can derive all the properties of $W_n$ algebra
in terms of the single function $\tau_n$.

Our method is so systematic that we can calculate the
transformation rules of the character polynomials without any
difficulity.
We write some of the useful  identities,
\begin{eqnarray}
[w_n(\zeta_1), \psi^*(\zeta_2)] & = & -\frac{1}{\zeta_1 - \zeta_2}
\frac{\partial^n}{\partial \zeta_2^n} \psi^*(\zeta_2),\nonumber\\
\psi^*(\zeta_1)\cdots\psi^*(\zeta_n)Q_{\ell}(w_0(\zeta)) & = &
Q_\ell(w_0+\sum_{s=1}^m\frac{1}{\zeta - \zeta_s})\psi^*(\zeta_1)
\cdots \psi^*(\zeta_m)\nonumber\\
Q_\ell(\sum_{s=1}^m \frac{1}{\zeta - \zeta_s}) & = &
\prod_{s=1}^{m}(\zeta - \zeta_s)^{-1}\partial^\ell_\zeta \prod_{s=1}^m
(\zeta - \zeta_s), \quad \mbox{etc...}\nonumber
\end{eqnarray}
For example, the transformation rule of  $W_\Omega(\zeta)$ reads,
\begin{equation}
\delta_s(\zeta_1) W_\Omega(\zeta_2) =
\tilde{\delta}_s(\zeta_1, \zeta_2) W_\Omega(\zeta_2),
\end{equation}
with,
\begin{eqnarray}
\tilde{\delta}_s(\zeta_1,\zeta_2) & = &
\sum_{\ell=0}^{s}~_sC_\ell \left(
\frac{1}{\ell+1}Q_{s-\ell}(j(\zeta_1))Q_{\ell +1}(\frac{d}{dj}(\zeta_1))
\right)_-\nonumber\\
   &    &
+\frac{~_sC_\ell}{\zeta_1 - \zeta_2}
\left( Q_{s-\ell}(j(\zeta_1))Q_\ell(\frac{d}{dj}
(\zeta_1)) - Q_{s-\ell}(\frac{d}{dj}(\zeta_1))
\frac{\partial^\ell}{\partial \zeta_2^\ell}\right)_-.
\end{eqnarray}
Generalizations to the higher polynomials are straightforward.

\section{Effective time coordinates}
As we have remarked before, only finite number of ``times"
are needed to describe the Grassmanian $Gr^{(n)}$.
The simplest situation is the one-particle case,
where we can simply replace $\frac{\partial}{\partial t_s}$
by $\frac{\partial^s}{\partial t_1^s}$.
For $n$-particle states,
the situation is more complicated.  Note that the $n$-particle
state can be represented as the $n$-th wedge product of the single
particle wave functions, $\Omega = f_1 \Lambda\cdots\Lambda f_n$.
Let us denote the differentiation on the $\alpha$-th wave function
as $\frac{\partial}{\partial t_s}(\alpha)$.
Then it is obvious that
\begin{equation}
\frac{\partial}{\partial t_s} =
\sum_{\alpha = 1}^n \frac{\partial}{\partial t_s}(\alpha)
= \sum_{\alpha=1}^n\left(\frac{\partial}{\partial t_1}(\alpha)\right)^s
\end{equation}
The problem reduces to finding algebraicaly independent basis
for the symmetric function of $n$-variables.
The desired basis is clearly $\Gamma_n = \{ \frac{\partial}{\partial t_\ell}
\}_{\ell = 1,2,\cdots,n}$. The explicit expression of
$\frac{\partial}{\partial t_{s+n}}$ in terms of $\Gamma$
is more involved.  However, it is fortunate that these derivatives
appear only through the combinations,
$Q_\ell (\frac{d}{dj}(\zeta)) = (V_n^*(\zeta))^{-1}\partial^\ell_\zeta
V_n^*(\zeta)$.
Hence we need to express only $V_n^*$
in terms of the basis $\Gamma_n$.  On $n$-particle states,
$t_0$ derivative is simply replaced by $n$.
( $J_0|\mbox{$n$-particle}> = n|\mbox{$n$-particle}>$.)
Hence,
\begin{eqnarray}
V^*_n & = & \zeta^n\exp\left(-\sum_{\ell =1}^\infty
\frac{1}{\ell}\zeta^{-\ell}\frac{\partial}{\partial
t_\ell}\right)\quad({\mbox{$n$-particle}})\nonumber\\
   & = & \zeta^n\prod_{\alpha = 1}^n \exp\left(
-\sum_{\ell =1}^\infty \frac{1}{\ell}\zeta^{-\ell}
\frac{\partial^\ell}{\partial t_1^\ell}(\alpha)\right)
\nonumber\\
 & = & \zeta^n\prod_{\alpha = 1}^n\left(
1-\frac{1}{\zeta}\frac{\partial}{\partial t_1}(\alpha)\right)
\end{eqnarray}
Here, there appear only up to $n$-th order symmetric polynomial
 in $V^*_n$.  They can be explicitly expressed in terms of
$\Gamma_n$.  Let us introduce polynomials $p_n([c])$ as follows,
\begin{equation}
\sum_{\ell = 0}^\infty p_\ell([c])\zeta^\ell = \exp \left(
\sum_{\ell = 1}^\infty c_\ell \zeta^\ell\right).
\end{equation}
For example, $p_0 =1$, $p_1 = c_1$, $p_2 = \frac{1}{2}c_1^2 + c_2,
\cdots$.  $p_n$ is written out of $c_1,\cdots,c_n$.
We claim that
\begin{equation}
V^*_n(\zeta) = \sum_{\ell =0}^n \zeta^{n-\ell}p_\ell([c])|_{
c_s = -\frac{1}{s}\frac{\partial}{\partial t_s}}.
\end{equation}

Note that $V^*_n$ is an $n$-th order polynomial in terms of
$\zeta$.  This proves explicitly that $Q_\ell(w_0(\zeta))|
\mbox{$n$-particle}> = 0$, if $\ell>n$.  This fact was mentioned
before.

\section{Discussions}
To summarize, the deformation of the $n$-th order differential equation
is understood as the analogue of the B\"acklund transformation in the
KP hierarchy.  Its transformation is written by using a kind of
boson-fermion correspondence.  The main difference between the two
theory is that the number of time coordinates is finite in our case.
The natural question is what is the analogue of the KP equation
itself.  The answer is straightforward.  Let us think the deformation
induced by the derivative of a higher coordinate $t_s$,
$\delta f = \frac{\partial}{\partial t_s}f$ in (1).
Since $\frac{\partial}{\partial t_s} = \left( \frac{\partial}{\partial
t_1}\right)^s$ on the one-particle state, we can take $ X =
\left(\frac{\partial}{\partial t_1}\right)^s$.
The deformation of ${\cal L}_n$ follows directly from (8),
\begin{equation}
\frac{\partial}{\partial t_s} {\cal L}_n
= {\cal B}_s{\cal L}_n, \qquad {\cal B}_s \equiv
- ({\cal L}_n \partial_1^s {\cal L}_n^{-1})^-.
\end{equation}
If we start from the UGM defined through $S$, we would get the
higher KdV equations,
\begin{equation}
\frac{\partial}{\partial \tilde{t}_s} {\cal L}_n =
[ A_s, {\cal L}_n], \qquad A_s = ({\cal L}_n^{s/n})^+.
\end{equation}
We should not confuse these two nonlinear equations.

Throughout this letter, we do not pay enough attention to the
$U(1)$-constraint (\ref{u1:const}).
In our framework, this constraint does not arise naturally
and we need to impose it by hand.
In terms of $\tau$-function, this constraint is given by
\begin{equation}
\partial_1 \tau([t])|_{t_2 = t_3 = \cdots = 0} = 0.
\end{equation}
Although it is easy to obtain known formulae in our language,
we have not clarified the geometrical origin of this constraint.

In the subsequent paper, we shall discuss the Hamiltonian
structure in terms of $\tau$-function and discuss the
free field realization.  It will be also interesting to discuss
the $N=2$ supersymmetric extension of $W_n$ symmetry\cite{IK}
because it is formulated almost parallel to the bosonic case.
\vskip 10mm

{\it Acknowledgements}:
The author would like to thank J.L. Gervais for the discussions about
the classical $W_n$-symmetry and the Toda field theory.
He is also obliged to V.Fateev for his important
comments and encouragements.

%\newpage

\end{document}